\documentclass[12pt,preprint]{aastex}

\slugcomment{Not to appear in Nonlearned J., 45.}

\shorttitle{Discovery of an Old Stellar System}
\shortauthors{Sakamoto \& Hasegawa}

\begin{document}

\title{Discovery of a Faint Old Stellar System at 150 kpc
}

\author{Sakamoto, T.}
\affil{Okayama Astrophysical Observatory, National Astronomical Observatory 
of Japan, Asakuchi, Okayama 719-0232, Japan}
\email{sakamtty@cc.nao.ac.jp}
\and
\author{Hasegawa, T.}
\affil{Gunma Astronomical Observatory, Agatsuma, Gunma 377-0702, Japan}
\email{hasegawa@astron.pref.gunma.jp}

\begin{abstract}
We report the detection of a faint old stellar system at
$(\alpha,\delta)=(194.29^\circ,~34.32^\circ)$ (SDSS J1257+3419),
based on the spatial distribution of
bright red-giant branch stars in the Sloan Digital Sky Survey Data 
Release 4.
SDSS J1257+3419 has a half-light radius of $38\pm 12$ pc and 
an absolute integrated $V$-magnitude of $M_V=-4.8^{+1.4}_{-1.0}$ mag 
at a heliocentric distance of $150\pm 15$ kpc. 
A comparison between SDSS J1257+3419 and known Galactic halo objects 
suggests that 
SDSS J1257+3419 is either (a) a faint and small dwarf galaxy or 
(b) a faint and widely extended globular cluster.
In the former case, SDSS J1257+3419 could represent an entity of 
a postulated subhalo of the Milky Way.
Further photometric and dynamical study of this stellar system is vital to 
discriminate these possibilities.
\end{abstract}

\keywords{Galaxy: dwarf galaxy --- Galaxy: globular cluster}

\section{Introduction}

The currently favored scenarios of hierarchical
galaxy formation in a cold dark-matter (CDM) universe advocate
that a larger galaxy was formed via a successive merging and accretion events 
of smaller galaxies like dwarf galaxies.
However, recent high-resolution numerical simulations based on CDM models have
revealed that the properties of CDM-based galaxies disagree with their 
properties observed
on a scale smaller than $\sim1$ Mpc (e.g., Navarro \& Steinmetz 2000).
One of the most serious discrepancies is the missing satellite-galaxy problem:
a considerably larger number of 
dark-matter clumps with the size of dwarf galaxy, or subhalos, 
are predicted to be distributed in the Galaxy-sized dark halo
than the observed dwarf galaxies in the Galaxy, 
particularly, at the low-mass end (Klypin et al. 1999; Moore et al. 1999).
This is partly due to the serious imcompleteness in the detection of 
Galactic dwarf galaxies.
Alternatively, no stars may be formed in large numbers of low-mass 
subhalos because star formation is efficiently suppressed by
the photoionoization from the first stars and active galactic nucleus
(Bullock, Kravtsov, \& Wenberg 2000).
Thus, the search for low-mass dwarf galaxies and the determination
of their stellar content is indispensable on
addressing the missing satellite-galaxy problem.

Sloan Digital Sky Survey (SDSS) is an ongoing deep spectroscopic and 
multicolor photometric survey.
Previous dwarf-galaxy searches based on the SDSS database 
mainly focused on the Galactic
dwarf galaxies with angular sizes of $\sim 10'$, 
where fluctuations on scales smaller than a few arcmin were smoothed and 
missed (Willman et al. 2002, 2005a; Zucker et al. 2006). 
However, Galactic 
dwarf galaxies with physical sizes of $\lesssim$ 100 pc at 200 kpc
and M31 dwarf galaxies (Zucker et al. 2004) have 
an angular size of a few arcmin, and then their dwarf galaxies 
may have still been missed.

Bright red-giant branch (RGB) stars are expected to be an effective probe to
search for distant stellar systems because they have a high intrinsic
luminosity.
In this paper, we search for stellar systems with sizes of a few arcmin
on the basis of the spatial distribution of the RGB stars in the
SDSS Data Release 4 (DR4),
and we report the detection of 
a faint old stellar system near the outskirt of the Galactic halo.

\section{Method and Results}

\subsection{Discovery of a faint stellar system}
DR4 provided us with deep imaging data that covered 6670 square degrees 
using five broadband filters ($u, g, r, i$, and $z$, 
Abazajian et al. 2004).
In the database, the objects with $r<21.5$ mag are confidently 
divided into stars and galaxies 
on the basis of their light profiles (Ivezic et al. 2000).

We constructed a color-selected sample of
bright RGB stars from the stars with $r<21.5$ mag in the DR4 database
such that they satisfy the following color conditions: 
$0.09+1.57(r-i)<g-r<0.59+1.57(r-i)$ and $0.4<g-r<1.2$ mag,
where the colors are corrected for reddening due to Galactic 
extinction (Schlegel, Finkbeiner, \& Davis 1998).
If the stars are giants, the selection 
corresponds to the stars with effective tempartures in the range of 
$3750<T_{eff}({\rm K})<5000$ and with metallicities 
in the range of $-5.0<{\rm [Fe/H]}<+1.0$, as shown in Lenz et al. (1998).
Note that the sample still includes dwarf stars with almost similar colors 
in Sloan's photometric system (Lenz et al. 1998).

The bright RGB stars were counted using a $5'\times 5'$ cell over the entire 
field covered by the DR4 database,
and then we searched for the spatial overdensities that were 
5 $\sigma$ and 10 stars above the mean background density.
The overdensities in the widely extended bright galaxies and their 
neighborhood were excluded, for the purpose of avoiding a serious 
incompleteness in the detection of stars in the galaxies 
(Mandelbaum et al. 2005). 
We verified the presence of RGB and horizontal branch (HB) in overdensities
in the color-magnitude diagram (CMD), 
which are the characterstic features of old stellar systems.
These checks were carried out in order to assess the contribution of dwarf 
star contamination (see preceding paragraph).

An unknown spatial overdensity is detected in a field centered
at $(\alpha,\delta)=(194^\circ.29,34^\circ.32)$ in addtion to 
known Galactic objects (i.e., dwarf galaxies, globular clusters, 
and an open cluster).
Figure 1(a) shows the spatial distribution of DR4's 
stars around this overdensity,
and they are symmetrically concentrated within $2'.5$.
In Figure 1(e), the CMD of the overdensity within a circle with a radius of 
$2'.5$ has a clear RGB at $0.4<g-r<0.8$ mag, $18.8<r\lesssim 23.0$ mag,
and a clear HB within $-0.35<g-r<-0.05$ mag, $21.5<r<22.0$ mag.
This is in contrast to the CMD of the same-sized control field 
at a distance of $1^\circ$  from the center of the overdensity in Figure 1(d).
We allocate the stars in a region enclosed within
$-10(g-r)+25.7<r<-10(g-r)+29.0$ and $18.8<r<23.0$ mag 
for RGB stars 
and within $-0.35<g-r<-0.05$ mag and $21.5<r<22.0$ mag for HB stars.
Both RGB and HB stars are strongly concentrated in Figures 1(b) and (c)
and their densities in a $5'\times 5'$ field centered at the overdensity
are $5\sigma$ and $9\sigma$ 
above the mean background stellar density, respectively.
Since these stars belong to old stellar populations, 
the overdensity is definitely a stellar system dominated by old stars
(hereafter, SDSS J1257+3419).

\subsection{Basic parameters of SDSS J1257+3419}
We estimate the basic parameters of SDSS J1257+3419, such as its
distance, metallicity, structural parameters, and total absolute magnitude.
First we estimate its distance and metallicity by referring to
the ridge lines of the RGB and HB stars in Galactic globular clusters. 
In order to produce the templates of the ridge lines, 
we perform photometry with Gunn filters 
for a sample of six globular clusters using the 65-cm telescope 
at the Gunma Astronomical Observatory.
The clusters, M5, M12, M15, M53, M56, and M92, are selected to include
a wide range of spectroscopically-measured metallicities ([Fe/H]=$-1.48\sim
-2.29$) and of HB morphological-types (the HB ratio$=0.31\sim 0.98$, 
which denotes the ratio of the difference in the number of 
blue and red HB stars to all HB stars.
Seeing condition was $1''.5\sim2''.5$ in $r$.
Very small color-term coefficients and stable airmass dependence 
on the photometric calibration suggest a good recovery
of the CMD of the program clusters;
the distance moduli which are measured 
by fitting Padova isochrones (Girardi et al. 
2004) to the RGB, are confirmed to be 
as accurate as 0.05 $-$ 0.1 mag with the published values.
Figure 1(f) shows the CMD of SDSS J1257+3419 and
the three representative ridge lines of the globular clusters.
The fit of the ridge lines to the HB places SDSS J1257+3419
at a distance modulus of $(m-M)_{0} =20.7 \pm 0.1 $, 
corresponding to 150 kpc with a 10\% accuracy.
We limit on the metallicity on the basis of the RGB and HB morphology.
The curvature of the RGB suggests
that metallicity of SDSS J1257+3419 can be infered 
coarsely within 0.5 dex for very old single stellar population:
its metallicity is lower than that of the metal-rich entity 
M5 ([Fe/H] = $-1.25$), and it is as low as that of the 
globular clusters M92 ([Fe/H] = $-2.25$);
fits to the CMD with a slightly metal-rich entity 
M53 ([Fe/H]=$-2.0$) is still acceptable. 
The HB ratio for SDSS J1257+3419 is 0.86, 
where blue and red HB stars are assumed to have $g'-r'<0.0$ mag 
and $g'-r'>0.0$ mag, respectively.
Based on [Fe/H] vs. the HB ratio for Galactic globular clusters (Zinn 1993),
we suggest that SDSS J1257+3419 has a 
metal-poor nature ([Fe/H]$\lesssim -1.5$).
Note that the ratio estimate is weak due to small number 
statistics and the failure to exclude field stars.
Thus, the value of [Fe/H] for SDSS J1257+3419 is less than $-1.5$ and 
could be as low as $-2.0$ or $-2.25$.

Next, we estimate the structural properties of SDSS J1257+3419.
We derive its centroid from the density-weighted 
first moment of the distribution,
and estimate the average ellipticity by using the three density-weighted second
moments (Stobie 1980).
For tracers of the radial profile, we adopt the stars with 
$g-r<0.8$ and $r<23$ mag, which effectively reduce the 
foreground stars at each radius.
Figure 2 shows the radial profile of the stars, where
the average densities with elliptical annuli are plotted
after subtracting a constant background level ($0.45 {~\rm arcmin^{-2}}$).
An exponential law with a half-light radius of $r_h=38$ pc approximately
represents the radial profile;
the radius little  depends on the binning size of the profile.

Finally, we roughly estimate an absolute integrated magnitude in the $V$-band, 
$M_V$, using two different methods.
This is because photometric data on faint RGB and bright main-sequence stars 
and kinematical data for membership identification are not available.
We first estimate its lower bound by summing the luminosities of the
RGB stars and HB stars within $r_h$ and then by doubling the summed luminosity
(e.g., Willman et al. 2005b).
The lower bound is estimated as $M_V=-3.4$ mag, 
assuming that the distance is 150 kpc.
We note that this value may be underestimated because of the highly 
distributed nature of bright RGB stars beyond the half-light radius 
in the stellar system.
We second estimate a rough value of $M_V$ from the number of cluster RGB stars 
that are brighter than the HB stars by applying
a loosely defined relation between the number of such RGB stars in the 
globular clusters and its $M_V$.
The relation is constructed by performing 
a weighted least-squares fitting procedure 
(weights being inversely proportional to the poisson noise of the number of 
the stars)
for the number of the observed RGB stars of
4 globular clusters (M5, M12, M52, and M92), which are expected to 
have small numbers of contamination and of missing RGB stars,
and  their values of $M_V$ corrected for the field of view (Harris 1996).
We find the most likely value of $M_V=-4.8$ mag and
the upper bound of $M_V=-5.8$ mag, which is $1\sigma$ above it.

The estimated basic parameters are listed in Table 1.

\section{Discussion}
We discuss whether SDSS J1257+3419 is a subhalo.
Although the direct way of the discrimination is 
the comparison between the total and stellar mass distributions,
the kinematical information to determine the total mass distribution 
is not available.
Previous works suggested that the Galactic globular clusters (GCs)
and dwarf galaxies (DGs) are 
dominated by stars and dark matter, respectively.
We thus set limits on the existence of the dark matter in SDSS J1257+3419,
by comparing it with GCs and DGs in some photometric and structural properties.

The DGs are relatively well separated from the GCs
in the $r_h$-$M_V$ plot shown in Figure 3.\footnote{We plot the
compilation data of Harris (1996) and Willman et al. (2005b) for the GCs
and that of Mateo (1998), Willman et al. (2005a), Zucker et al. 2006),
and Belokurov et al. (2006) for the DGs.}
The half-light radius of SDSS J1257+3419 is
larger than the GCs and to be smaller than the DGs.
SDSS J1257+3419 also has a luminosity comparable to the
faintest GCs and DGs.
Thus, SDSS J1257+3419 is located near the boundary between the
dwarf galaxies and globular clusters:
it could be a faint, small dwarf galaxy, 
or a faint, extended globular cluster.

It should be noted that 
another stellar system having such intermediate properties---SDSS J1049+5103---
has been recently detected (Willman et al. 2005).
The follow-up observations showed 
that SDSS J1049+5103 is a stellar system with multiple stellar tails at
a distance of $\sim40$ kpc 
and that its large spatial extent relative to low-luminosity globular 
clusters may be formed as a result of tidal interaction with the Galaxy
during their orbital motions (Willman et al. 2006).
SDSS J1257+3419 is more luminous and more distant than SDSS J1049+5103 
($M_V=-2.5$ mag),
and the mass-loss in the former system due to Galactic tides  
might not be sufficiently high to explain its large spatial extent.
In order to determine whether it is a dwarf galaxy or a globular cluster,
it is vital to determine the photometric and structural properties 
from a large sample of stars down to the turnoff of the main sequence
and to determine the mass distribution from their kinematical information.

Further we discuss the properties of SDSS J1257+3419.
If SDSS J1257+3419 is Galactic DG,
it is the fifth furthest one.
Thus, it is a good tracer of the mass determination of
the Galaxy (Sakamoto et al. 2003).

If SDSS J1257+3419 is a GC, it is the furthest one.
It might not be gravitationally bound to the Milky Way,
but to the Local Group.
Its shape is almost spherical, and it might have a simple dynamical evolution
due to weak Galactic tides.
Thus, it might be a good entity to understand the evolution of an isolated GC.

We set limits on the star formation history. 
The known Galactic dSphs exhibit a wide variety of star formation histories
ranging from galaxies that are entirely dominated by very old populations to 
dwarf galaxies with considerable fractions of intermediate-age stars 
(Mateo 1998; Grebel \& Gallagher 2004). 
Assuming no star within $22<r<23$ mag and $g-r<0.4$ mag (see Figure 1(e)), 
we find that no star formation occured at least during the last four Gyr.
The absence of a red clump in the CMD strengthen that no sizable fraction of intermediate-age stars are populated in the SDSS J1257+3419.


\section{Conclusion}
We detect a faint old stellar system (SDSS J1257+3419)
with a half-light radius of 38 pc 
at a heliocentric distance of $150\pm 15$ kpc on the basis of the spatial 
distribution of bright RGB stars in SDSS-DR4.
The rough absolute integrated $V$-magnitude is $-4.8^{+1.4}_{-1.0}$ mag.
When we compared the half-light radius and absolute integrated magnitude with 
those of the 
known Galactic objects (i.e., dwarf galaxies and globular clusters), 
we find that SDSS J1257+3419 is either a faint and small dwarf galaxy or
a faint and widely extended globular cluster.

\acknowledgments
TS would like to thank
Okayama Prefecture and the National Institute of Information
and Communications Technology for their support on high-speed network 
connection (Okayama Information Highway and Japan Gigabit Network) for
data transfer and analysis.
TH thanks Y. Kondo and Y. Negishi for their assistance in ridge line
measurements in this study.

Funding for the Sloan Digital Sky Survey (SDSS) has been provided by the Alfred P. Sloan Foundation, the Participating Institutions, the National Aeronautics and Space Administration, the National Science Foundation, the U.S. Department of Energy, the Japanese Monbukagakusho, and the Max Planck Society. The SDSS Web site is http://www.sdss.org/.
The SDSS is managed by the Astrophysical Research Consortium (ARC) for the Participating Institutions. The Participating Institutions are The University of Chicago, Fermilab, the Institute for Advanced Study, the Japan Participation Group, The Johns Hopkins University, the Korean Scientist Group, Los Alamos National Laboratory, the Max-Planck-Institute for Astronomy (MPIA), the Max-Planck-Institute for Astrophysics (MPA), New Mexico State University, University of Pittsburgh, University of Portsmouth, Princeton University, the United States Naval Observatory, and the University of Washington.

\clearpage

\begin{figure*}
\epsscale{1.0}
\plotone{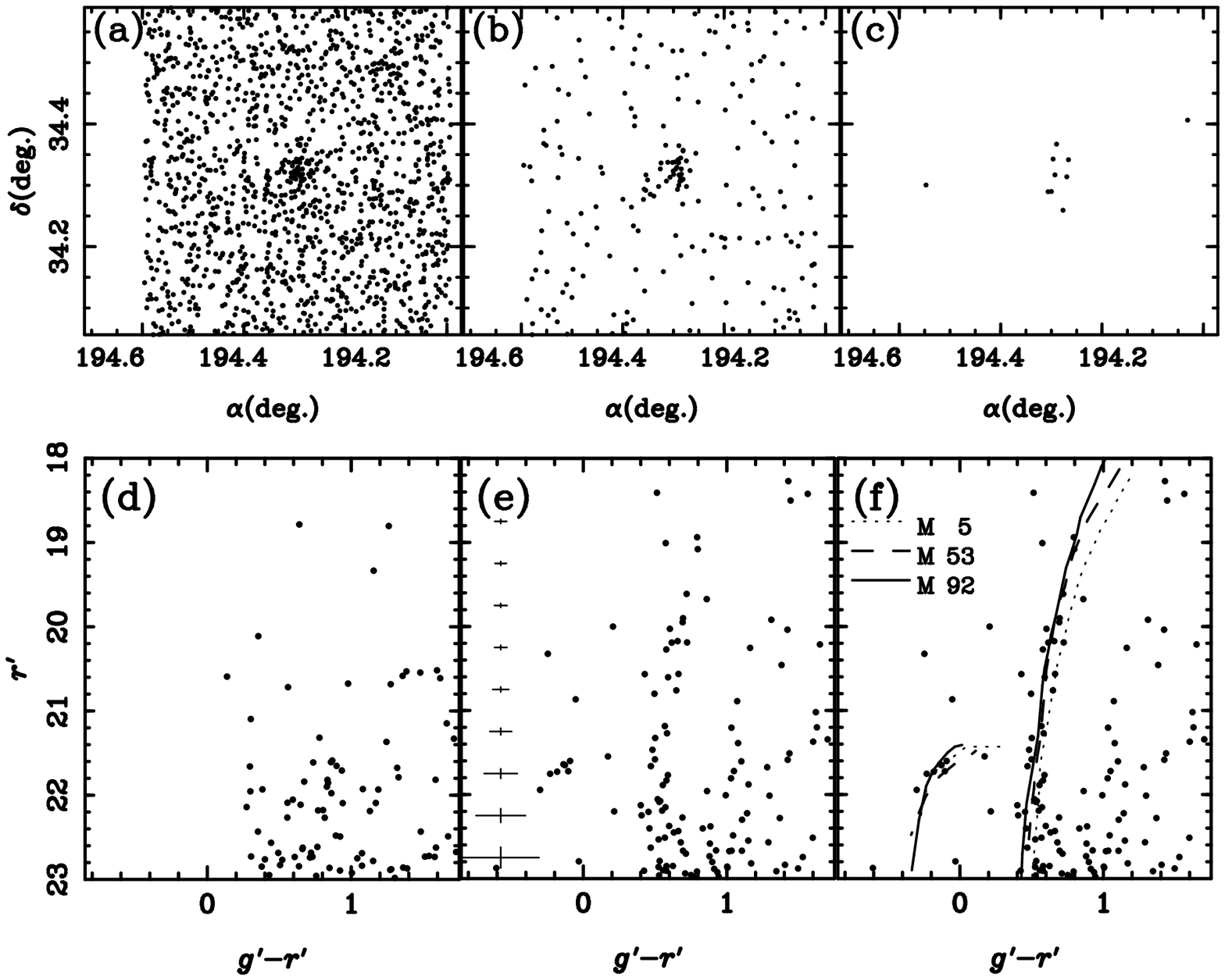}
\caption{Spatial distribution and CMD for SDSS J1257+3419.
Upper panels show the spatial distribution around 
SDSS J1257+3419 for (a) all the stars, (b) RGB stars, 
and (c) HB stars.
Lower panels show the CMD in
a $2'.5$-radius circle centered on SDSS J1257+3419 (panels e and f) 
and in the same-sized control field at a distance of $1^\circ$ from its center
(panel d). 
The curves in panel f denote the ridge lines of three globular clusters.
}
\label{fig1}
\end{figure*}

\clearpage
\begin{figure}
\epsscale{.80}
\plotone{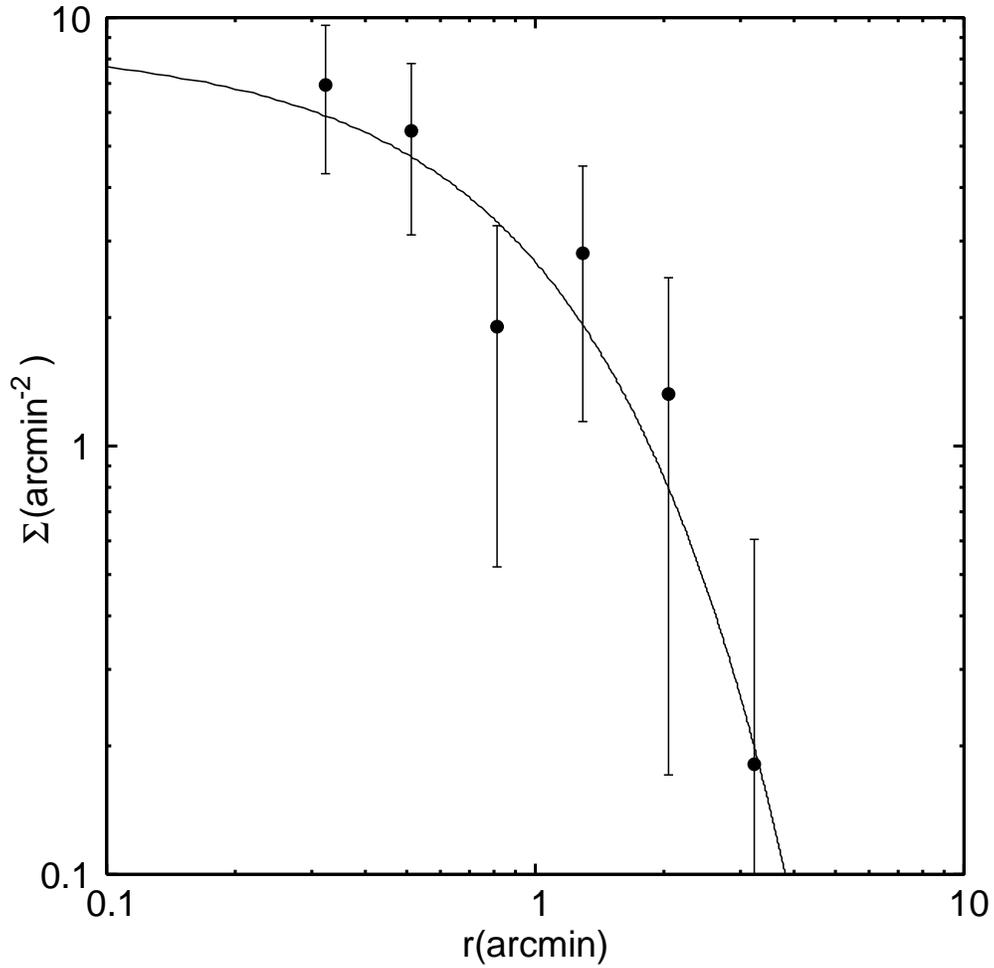}
\caption{Radial profile of SDSS J1257+3419 for the stars with $g-r<0.8$ mag 
and $r<23$ mag.
The surface number densities, $\Sigma$, denote 
the average densities within the elliptical annuli 
after subtracting a constant background level ($0.45 {~\rm arcmin^{-2}}$).
The solid curve denote the best-fitting exponential 
law with a half-light radius
of $0.86'$.}
\label{fig2}
\end{figure}

\begin{figure}
\epsscale{0.80}
\plotone{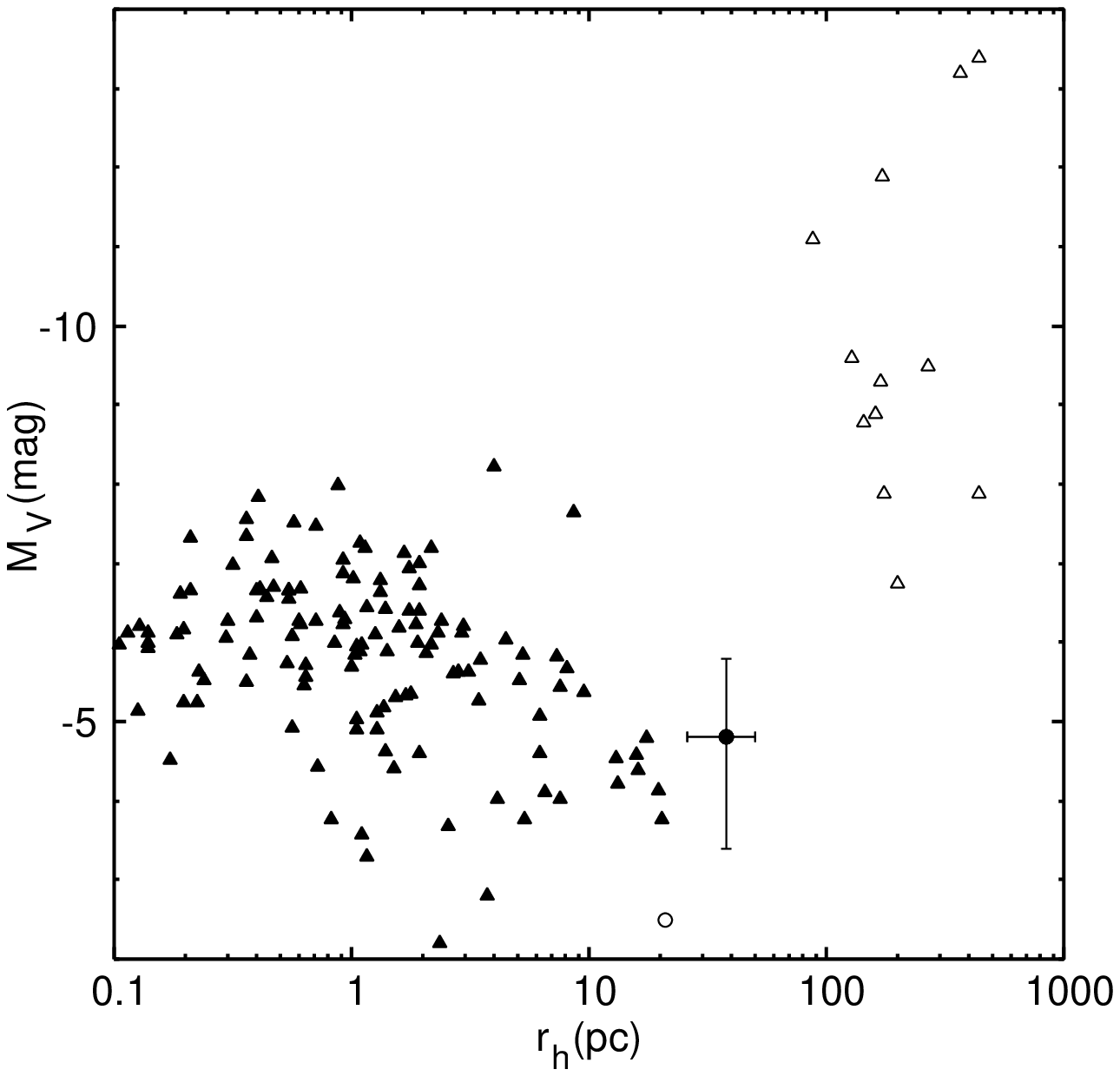}
\caption{Plot of half-light radius, $r_h$, and 
absolute integrated $V$-magnitude, $M_V$, 
for SDSS J1257+3419 (filled circle), SDSS J1049+5103 (open circle), 
Galactic globular clusters (filled triangles), and 
dwarf galaxies (open triangles).}
\label{fig2}
\end{figure}


\begin{table}
\begin{center}
\caption{Basic parameters of SDSS J1257+3419}
\begin{tabular}{ll}
\tableline\tableline
Parameter&Value\\
\tableline
Coordinates (J2000)&$(12^h57^m09.6^s$, $+34^\circ 19'12'')$\\
Galactic coordinates&$(113.59^\circ,82.70^\circ)$\\
Ellipticity&0.10\\
$A_V$&0.00 mag\\
$(m-M)_0$&$20.7\pm 0.1$ mag\\
Heliocentric distance&$150\pm 15$ kpc\\
$r_h$&$38 \pm 12$ pc\\
$M_{tot,V}$&$-4.8^{+1.4}_{-1.0}$ mag\\
\tableline
\end{tabular}
\end{center}
\end{table}

\end{document}